# Layer thickness dependence of the current induced effective field vector in Ta|CoFeB|MgO

Junyeon Kim[1], Jaivardhan Sinha[1], Masamitsu Hayashi[1*], Michihiko Yamanouchi[2,3], Shunsuke Fukami[2], Tetsuhiro Suzuki[4], Seiji Mitani[1] and Hideo Ohno[2,3,5]

[1]*National Institute for Materials Science, Tsukuba 305-0047, Japan*
[2] *Center for Spintronics Integrated Systems, Tohoku University, Sendai 980-8577, Japan*
[3]*Laboratory for Nanoelectronics and Spintronics, Research Institute of Electrical Communication, Tohoku University, Sendai 980-8577, Japan*
[4]*Renesas Electronics Corporation, Sagamihara 252-5298, Japan*
[5]*WPI Advanced Institute for Materials Research, Tohoku University, Sendai 980-8577, Japan*

The role of current induced effective magnetic field[1] in ultrathin magnetic heterostructures is increasingly gaining interest since it can provide efficient ways of manipulating magnetization electrically[2-3]. Two effects, known as the Rashba spin orbit field[4] and the spin Hall spin torque[3], have been reported to be responsible for the generation of the effective field. However, quantitative understanding of the effective field, including its direction with respect to the current flow, is lacking. Here we show vector measurements of the current induced effective field in Ta|CoFeB|MgO heterostructrures. The effective field shows significant dependence on the Ta and CoFeB layers' thickness. In particular, 1 nm thickness variation of the Ta layer can result in nearly two orders of magnitude difference in the effective field. Moreover, its sign changes when the Ta layer thickness is reduced, indicating that there are two competing effects that contribute to the effective field. The relative size of the effective field vector components, directed transverse and parallel to the current flow, varies as the Ta thickness is changed. Our results illustrate the profound characteristics of just a few atomic layer thick metals and their influence on magnetization dynamics.

*Email: hayashi.masamitsu@nims.go.jp

Spin orbit coupling plays an important role in ultrathin magnetic heterostructures. It has been reported that the Rashba effect[5], present in systems with large spin orbit coupling and structure inversion asymmetry, can generate significant amount of current induced effective magnetic field[6-9] which, for example, enables fast domain wall motion in ultrathin Co layer sandwiched by Pt and AlO$_X$[4]. Similarly, spin orbit coupling in heavy metals can generate spin current via the spin Hall effect[10], which can also develop effective field in its neighboring magnetic layer[11-12]. Using the giant spin Hall effect of Ta, power efficient magnetization switching of the adjacent CoFeB layer has been demonstrated[3].

However, the size and direction of the current induced effective field seems to vary depending on the system and the underlying mechanism of such field generation is not well understood. For example, the effective field in Pt|Co|AlOx is reported to be ~3000 to ~10000 Oe for a current density of $10^8$ A/cm$^2$, pointing perpendicular to both the film normal and the current flow direction (defined as a transverse field hereafter)[4,13]. More recently, in the same system, signs of current induced effective field directed along the current flow, i.e. the "longitudinal field", have been observed[14]. This longitudinal field has been associated with either the combination of the Rashba effect and spin accumulation[15] or the spin Hall effect from the Pt layer[16-19]. Similarly, in Ta|CoFeB|MgO, it has been reported that a giant spin Hall effect in Ta, although its sign is opposite to that of Pt, generates the longitudinal effective field[3]. In Ta|CoFeB|MgO, however, contradictory results have been reported for the transverse effective field; on one hand, transverse field of ~1900 Oe at $10^8$ A/cm$^2$ was observed[20] with its direction opposite to that of Pt|Co|AlO$_X$, whereas on the other, no indication of such field was found[3]. Given the broad interest of Ta|CoFeB|MgO heterostructures for possible applications including random access memories[21-22] and domain wall based devices[23], it is essential to reveal the underlying physics of current induced effects.

Here we show a systematic study of the current induced effective field in Ta|CoFeB|MgO. We use a low current excitation technique to quantitatively evaluate the size and direction of the effective field. We find that the size and even the sign of the transverse and longitudinal effective fields vary as the Ta layer thickness is changed, suggesting competing contributions from two distinct sources. We find that the transverse effective field is larger than the longitudinal field, by nearly a factor of three, when the Ta layer thickness is large. In contrast, the relative size of the two components shows somewhat an oscillatory dependence on the Ta thickness for films with thin Ta.

Films are sputtered on a highly resistive silicon substrate coated with 100 nm thick thermally oxidized $SiO_2$. We use a linear shutter during the sputtering to vary the thickness of one layer in each substrate. Two film stacks are prepared here: Ta wedge: Si-sub|$d_{Ta}$ Ta|1 $Co_{20}Fe_{60}B_{20}$|2 MgO|1 Ta and CoFeB wedge: Si-sub|1 Ta|$t_{CoFeB}$ $Co_{20}Fe_{60}B_{20}$|2 MgO|1 Ta (all units are in nanometer). Note that the composition ( in atomic percent) of the CoFeB is different from our previous studies which were mostly $Co_{40}Fe_{40}B_{20}$[20,23-24]. We set the thickness of each layer, $d_{Ta}$ and $t_{CoFeB}$, to vary from ~0 to ~2 nm. Throughout this paper, the nominal thickness is used for the Ta wedge. For the CoFeB wedge film, correction of the thickness was required due to technical reasons, and thus we use our resistivity results to calibrate the thickness. All films are annealed at 300 °C for one hour ex-situ after the film deposition. Photo-lithography and Ar ion etching are used to pattern Hall bars from the film and a lift off process is used to form the contact electrodes (10 Ta|100 Au). Prior to the deposition of the contact electrodes, we etch the Ta capping layer and nearly half of the MgO layer to avoid large contact resistance. Although etching of the MgO layer significantly influences the magnetic anisotropy of the CoFeB layer under the etched region[24], here we assume that this has little effect on the evaluation of the current induced effective fields since we limit the applied field smaller than the magnetization switching field.

Schematic illustration of the experimental set up and definition of the coordinate system are shown in Fig 1(a). The width and length of typical wires measured are 10 μm and 60 μm, respectively. We measure wires with different width, ranging from 5 μm to 30 μm, and find little dependence on the width for most of the parameters shown here. Positive current is defined as current flowing along the +y direction in Fig 1(a). Current is fed into the wire and the Hall voltage is measured in all experiments. Using the Extraordinary Hall Effect (EHE), we study the magnetic and transport characteristics of each device.

Fig. 1(b) and 1(c) show the wedge layer thickness dependence of $\Delta R_{XY}$, defined as the difference in the EHE resistance when the magnetization of the wire is pointing along +z and –z. Here we apply constant current $I_{DC}$ (10 or -10 uA) and measure the Hall voltage $V_{XY}$; $R_{XY}= V_{XY}/I_{DC}$. For the Ta wedge, $\Delta R_{XY}$ decreases as the Ta thickness is increased, whereas it increases with the CoFeB thickness for the CoFeB wedge. This is primarily due to the change in the amount of the current that flows into the CoFeB layer; the larger the current that flows in the CoFeB layer, the larger the $\Delta R_{XY}$. For the CoFeB wedge, $\Delta R_{XY}$ tends to saturate at ~1.2 nm, which may indicate that the *intrinsic* $\Delta R_{XY}$ (where no current flow is assumed other than the CoFeB layer) starts to drop for thicker CoFeB.

The coercivity of the wire is plotted in Fig. 1(d) and 1(e) as a function of the wedge layer thickness. The EHE resistance is monitored during an out of plane field sweep and we measure the switching field $H_{SW}^+$ and $H_{SW}^-$ for ascending and descending field sweeps, respectively. The coercivity $H_C$ is defined as $H_C =(H_{SW}^+ - H_{SW}^-)/2$. The variation of the coercivity with the wedge layer thickness is rather scattered from wire to wire, especially for the Ta wedge. The variation can be due to small changes in the degree of etching of the MgO layer, which can influence the coercivity[24]. Note that we do not find any systematic dependence of $H_C$ on the wire width. Overall, the coercivity tends to increase as the Ta thickness is reduced or the CoFeB thickness is increased. This is in accordance with the thickness dependence of the perpendicular magnetic anisotropy, that is, the higher the anisotropy, the higher the $H_C$.

The current induced effective field is measured using a low current excitation lock-in technique[13]. We apply a constant amplitude sinusoidal voltage to the wire and measure the Hall voltage with a lock-in amplifier. The in-phase first harmonic ($V_\omega$) and the out of phase (90 deg off) second harmonic ($V_{2\omega}$) signals are measured simultaneously using two independent lock-in amplifiers. We sweep the in-plane field directed transverse ($H_T$) or parallel ($H_L$) to the current flow to obtain the transverse and the longitudinal components of the effective field vector. The transverse ($\Delta H_T$) and longitudinal ($\Delta H_T$) effective fields are obtained by the following equations (see supplementary material S1[25]).

$$\Delta H_{T(L)} = -2 \frac{\partial V_{2\omega}}{\partial H_{T(L)}} \bigg/ \frac{\partial^2 V_\omega}{\partial H_{T(L)}^2} \qquad (1)$$

Fig. 2 shows the in-plane field dependence of $V_{2\omega}$ when a 20 V peak to peak AC voltage is applied to the wire. The transverse (Fig. 2(a) and 2(c)) and the longitudinal (Fig. 2(b) and 2(d)) field sweeps are shown for two wires with different Ta thicknesses (~0.3 and ~1.2 nm). The corresponding field dependence of the first harmonic signal $V_\omega$ shows similar structure for all four cases (see supplementary material S2). However the second harmonic signal $V_{2\omega}$ depends on the applied field direction as well as the film structure. For longitudinal field, the slopes of $V_{2\omega}$ versus the field are the same for both magnetic states pointing along +z or –z, whereas their sign reverse for the transverse field. Interestingly, the sign of the slope changes for both longitudinal and transverse applied fields when the Ta thickness is varied from the thin limit to the thicker side. Since the curvature of $V_\omega$ (ι.ε. $\partial^2 V_\omega / \partial H_{T(L)}^2$ ) hardly changes for each condition, these results indicate that the effective field changes depending on the applied field direction as well as the film structure.

We plot $\Delta H_T$, deduced from Eq. (1), as a function of the input voltage $V_{IN}$, corresponding to the peak to peak value of the AC excitation voltage, in Fig. 3 for the Ta (Fig. 3(a)) and CoFeB (Fig. 3(b)) wedge films. The corresponding current density for $V_{IN}$=1 V is ~3.4-4.1x10$^5$ A/cm$^2$ if we assume uniform current flow across the Ta underlayer and the CoFeB layer. The variation in the current density is due to the change in the contact resistance, which includes resistance from the ~1 nm thick MgO layer, since two-point probe resistance measurements are used. Data from the four-point probe resistance measurements are shown in supplementary information S3. In Fig. 3(a) and 3(b), $\Delta H_T$ is plotted for both magnetization states pointing along +z and –z. We find that $\Delta H_T$ is independent of the magnetization state and shows nearly a linear dependence on $V_{IN}$. Note that the maximum $V_{IN}$ is limited to a certain value for each wire due to the reduction in the switching field by the application of current. The reduction of the switching field tends to occur at lower $V_{IN}$ for wires with thicker Ta and/or thinner CoFeB layers.

The slope of $\Delta H_T$ versus $V_{IN}$ at low $V_{IN}$, as shown by the solid lines, is plotted as a function of the Ta and CoFeB layer thickness in Fig. 3(c) and 3(d), respectively. For the Ta wedge sample, $\Delta H_T/V_{IN}$ shows a significant variation with the Ta layer thickness: it changes by nearly two orders of magnitude and even changes its sign when the thickness is ~0.6 nm. Note that the positive sign of the transverse effective field $\Delta H_T/V_{IN}$ agrees with previous study of Ta|CoFeB|MgO [20] and is opposite to that of Pt|Co|AlO$_X$ [4,13]. The change in $\Delta H_T/V_{IN}$ with the CoFeB layer thickness is also significant; however, it monotonically decreases with increasing the thickness.

The longitudinal effective field, $\Delta H_L$, is plotted as a function of $V_{IN}$ in Fig. 4 for the Ta (Fig. 4(a)) and CoFeB (Fig. 4(b)) wedge films. For both films, the sign of $\Delta H_L$ depends on the magnetization direction. This shows that the longitudinal effective field, which is orthogonal to $\Delta H_T$, takes the form of $\vec{m} \times \Delta \vec{H}_T$, where $\vec{m}$ is a unit vector representing the magnetization direction. As with $\Delta H_T$, $\Delta H_L$ scales linearly with $V_{IN}$. The slope of $\Delta H_L$ versus $V_{IN}$, $\Delta H_L/V_{IN}$, is plotted as a function of Ta and CoFeB layer thickness in Figs. 4(c) and 4(d), respectively. The longitudinal effective field $\Delta H_L/V_{IN}$ decreases with decreasing Ta thickness and it changes its sign when the thickness is below ~0.5 nm. In contrast, $\Delta H_L/V_{IN}$ shows almost no dependence on the CoFeB layer thickness.

To directly show the relative magnitude of the transverse and longitudinal fields, we plot the ratio of the two, defined as $R=[\Delta H_L*sgn(-M_Z)]/\Delta H_T$, in Fig. 5(a) and 5(b) for the Ta and CoFeB wedge films, respectively. Since $\Delta H_L$ depends on the magnetization direction, we multiply the sign of $-M_Z$ ($\equiv sgn(-M_Z)$) to exclude the effect of the magnetization direction (the minus sign in front of $M_Z$ is to comply with the convention taken by recent theoretical works[7,12]). We first note that $|R|$ is less than 1 for most wires, which corresponds to transverse effective field being larger than the longitudinal field. We color code both figures to illustrate the direction (sign) of $\Delta H_T/V_{IN}$ and $\Delta H_L/V_{IN} \times sgn(-M_Z)$: blue is for positive and red is for negative values. Note that $\Delta H_T/V_{IN}$ is negative while $\Delta H_L/V_{IN} \times sgn(-M_Z)$ is positive when the Ta thickness is ~0.6 nm, which is why we find a negative R (color coded by purple).

Our results can be understood under the framework developed by Manchon *et al*[6,12]. According to their calculations, the effective field can be described by combination of two independent effects, the spin Hall spin torque[3,11-12] and the Rashba spin-orbit torque[6-9]. For both transverse and longitudinal fields, the spin Hall spin torque increases with the Ta layer thickness in the form that is proportional to $1-\mathrm{sech}(d_{Ta}/\lambda)$[3,19], where $\lambda$ is the spin diffusion constant of Ta. In contrast, the Rashba spin-orbit field is nearly independent of the Ta layer thickness. Our data can be qualitatively explained based on their calculations. The sign of the spin Hall current from previous reports[3,26] indicates that the effective field due to the spin Hall spin torque results in positive $\Delta H_L/V_{IN} \times \mathrm{sgn}(-M_Z)$, and positive $\Delta H_T/V_{IN}$ if any transverse component exists. Indeed, the effective field increases with the Ta thickness in this regime. On the other hand, the direction of the Rashba spin orbit field due to the magnetic|oxide layers interface is expected to point along the -x direction when a positive current is passed along the wire[4], which suggests negative $\Delta H_T/V_{IN}$ (and $\Delta H_L/V_{IN} \times \mathrm{sgn}(-M_Z)$ if it exists). Although the magnitude is much smaller, this more or less agrees with our observation where we find relatively small dependence on the Ta layer thickness for negative $\Delta H_T/V_{IN}$ for thin Ta film stacks. The sign change of the effective field thus likely represents the change in the dominant torque, whether it is the spin Hall or the Rashba spin-orbit torque. From Fig. 3(c) and 4(c), we infer that the transverse and longitudinal fields are dominated by the spin Hall spin torque when $d_{Ta}$ is larger than ~0.6 nm and ~0.5 nm, respectively.

It is interesting to note that both the transverse and the longitudinal effective fields scale to zero to a non-zero Ta thickness ($t_d$) when extrapolated from thicker Ta and taking into account the negative $\Delta H_T/V_{IN}$ and $\Delta H_L/V_{IN} \times \mathrm{sgn}(-M_Z)$ for thinner Ta: $t_d$~0.8 nm for the transverse and ~0.7 nm for the longitudinal effective field. This indicates that a certain Ta thickness is required to generate non-zero effective field. If we assume the source of the effective field for films with thick Ta underlayer is due to the spin Hall spin torque, then the presence of such threshold thickness may be due to a non-transparent Ta|CoFeB interface for passing spin currents across, or something more intrinsic related to the generation of the spin Hall current in Ta.

As shown in Fig. 5(a), the transverse effective field is nearly three times larger than the longitudinal component for the thick Ta regime, i.e. R~0.3. This implies that, if the effective field is due to the spin Hall current from the Ta layer, the injected spin current exert torque on the CoFeB layer in a way similar to the "field-like torque" proposed in magnetic tunnel junctions[27-29]. However, it is not clear what determines the ratio R between the transverse and longitudinal components for the spin Hall spin torque. For the Rashba spin orbit field, it has been reported[7] recently that the ratio should be given by the non-adiabatic spin torque term, typically represented by $\beta$. As shown in Fig. 5(a), we find that R is increasing toward 1 as the Ta thickness is decreased. Although the damping constant ($\alpha$) may depend on the Ta layer thickness, these results indicate a large non-adiabatic spin torque parameter for thin Ta film stacks, which has also been reported in other systems with large Rashba spin orbit fields ($\beta/\alpha \sim 1$)[30].

**Acknowledgements**

The authors thank helpful discussions with H-W. Lee and K-J. Lee. This work was partly supported by the Japan Society for the Promotion of Science (JSPS) though its "Funding program for world-leading innovate R & D on science and technology (FIRST program)".

**Figure captions**

**Fig. 1. Experimental setup and the magnetic properties of the Ta and CoFeB wedge films.** (a) Schematic image of the experimental set up. (b, c) $\Delta R_{XY}$ as a function of the Ta (b) and CoFeB (c) layer thickness. (d,e) Ta (d) and CoFeB (e) layer thickness dependence of the coercivity.

**Fig. 2. Second harmonic signals illustrating the variation in the effective field under different conditions.** (a-d) 90 deg out of phase second harmonic signal $V_{2\omega}$ plotted as a function of in-plane field directed transverse (a, c) and parallel to (b, d) to current flow for $V_{IN}$=20 $V_{pp}$. Solid and open symbols represent signals when the magnetization is pointing along +z and –z, respectively. The wedge layer thickness $d_{Ta}$ is 0.3 nm for (a, b) and 1.2 nm for (c, d).

**Fig. 3. Ta and CoFeB thickness dependence of the transverse effective field.** (a,b) Input voltage ($V_{IN}$) dependence of the transverse effective field $\Delta H_T$ for the Ta (a) and CoFeB (b) wedge films. Solid lines show linear fitting to the data. (c,d) Slope of $\Delta H_T$ versus $V_{IN}$ plotted as a function of Ta (c) and CoFeB (d) thickness. (c) Inset: magnification of the main plot showing the negative $\Delta H_T/V_{IN}$ at Ta thicknesses below ~0.6 nm. Solid and open symbol correspond to magnetization pointing along +z and –z, respectively.

**Fig. 4. Ta and CoFeB thickness dependence of the longitudinal effective field.** (a,b) Input voltage ($V_{IN}$) dependence of the longitudinal effective field $\Delta H_L$ for the Ta (a) and CoFeB (b) wedge films. Solid lines show linear fitting to the data. (c,d) Slope of $\Delta H_L$ versus $V_{IN}$ plotted as a function of Ta (c) and CoFeB (d) thickness. (c) Inset: magnification of the main

plot showing the sign reversal of $\Delta H_L/V_{IN}$ at Ta thicknesses below ~0.5 nm.   Solid and open symbol corresponds to magnetization pointing along +z and –z, respectively.

**Fig. 5. Ratio between the longitudinal and transverse effective fields.**   (a,b) The ratio, defined as $R=\Delta H_L \times sgn(-M_Z)/\Delta H_T$, is plotted as a function of Ta (a) and CoFeB (b) thicknesses. The background color represents the thickness range where $\Delta H_T/V_{IN}$ and $\Delta H_L/V_{IN} \times sgn(-M_Z)$ are positive (blue) or negative (red).   Purple indicates that $\Delta H_T/V_{IN}$ is negative whereas $\Delta H_L/V_{IN} \times sgn(-M_Z)$ is positive.

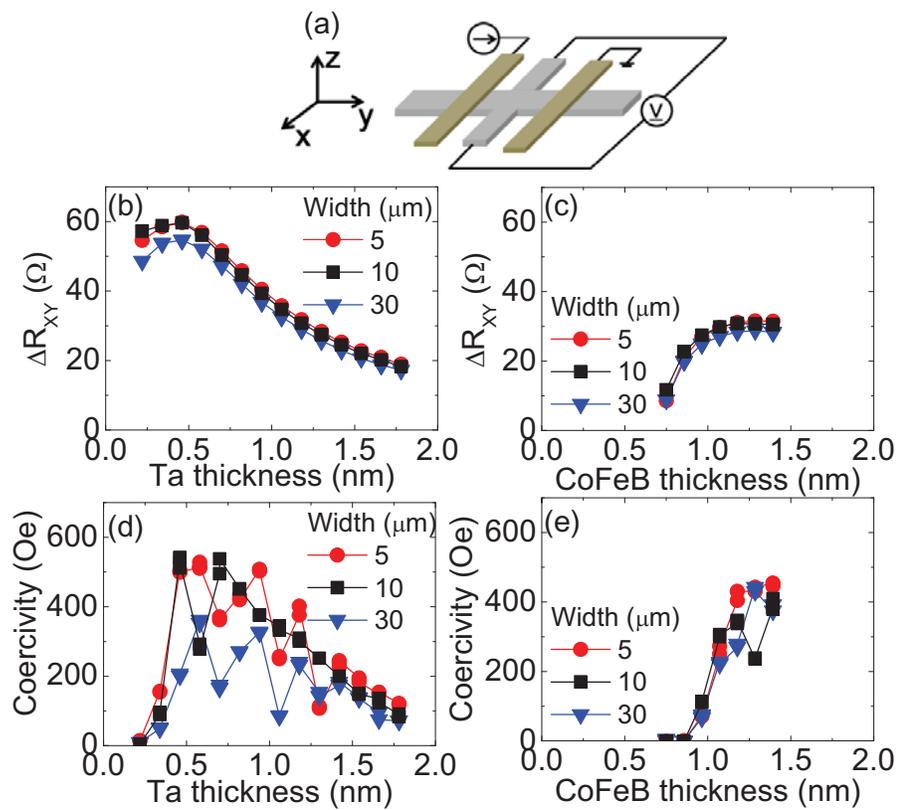

Fig. 1

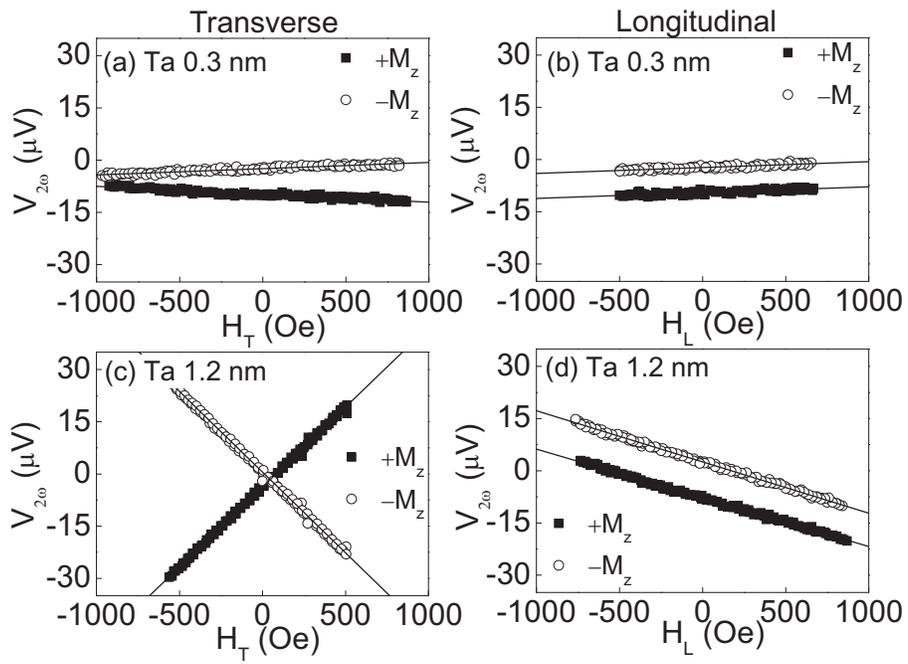

Fig. 2

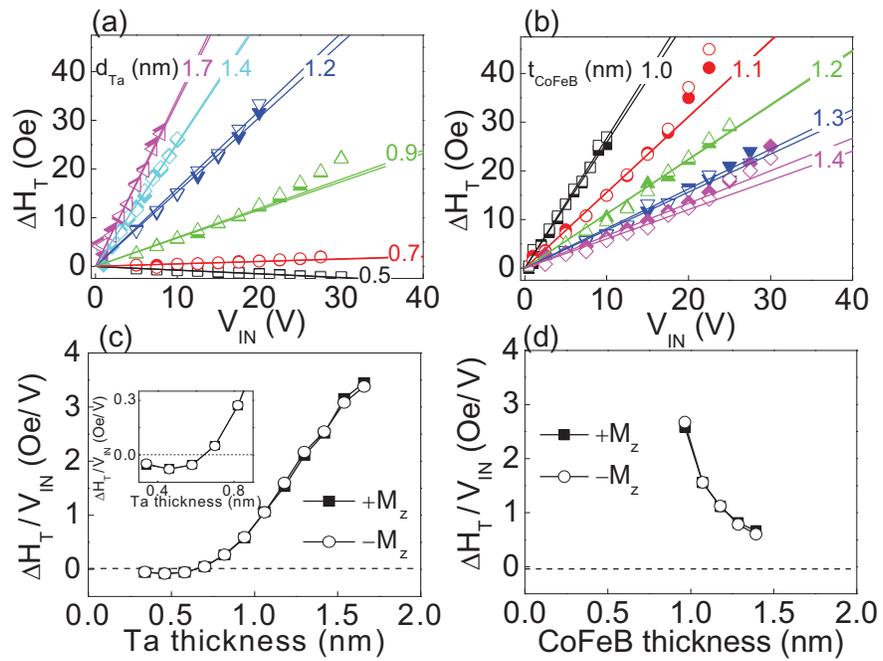

Fig. 3

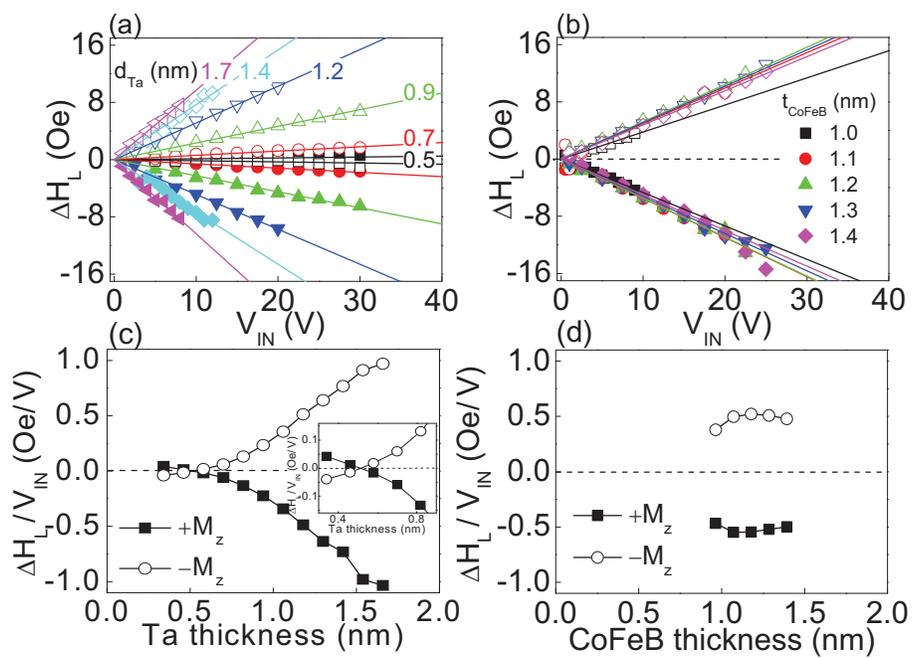

Fig. 4

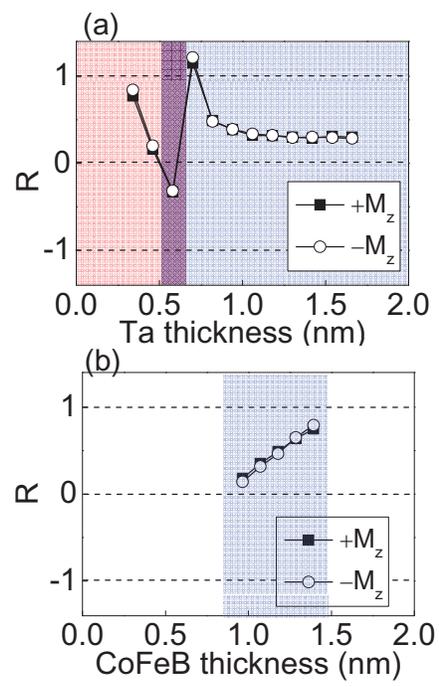

Fig. 5

## Supplementary information

### S1. Derivation of Eq. (1)

The magnetic energy of our system under consideration can be expressed as

$$E = \left(-K_u + 2\pi M_s^2\right)\cos^2\theta - M_s\left(H_T \cos\varphi \sin\theta + H_L \sin\varphi \sin\theta + H_z \cos\theta\right) \quad \text{(S1.1)}$$

where $K_u$ is anisotropy constant (positive for out of plane magnetization), $M_s$ is saturation magnetization, $H_z$, $H_T$, $H_L$ are the out of plane, in-plane transverse and longitudinal (along the current flow) fields, respectively. $\theta$ is the polar angle (with respect to the z axis) and $\varphi$ is the azimuthal angle (with respect to the x axis) of the magnetization. Assuming that the current induced effective field and the external in-plane field is small enough so that the magnetization tilt angle can be considered small, i.e. $\theta \ll 1$, the equilibrium tilt angle $\theta_0$ with respect the external field can be expressed as

$$\theta_0 = \frac{H_T \cos\varphi + H_L \sin\varphi}{D} \quad \text{(S1.2)}$$

where $D \equiv (2K_u/M_s) - 4\pi M_s + H_z$. If $H_z$ is set to zero, which is close to our experimental condition, D can be considered as a constant.

The current applied to the wire can be expressed in terms of the wire resistance $R_{XX}$ and the AC voltage with amplitude $V_{IN}$ and frequency $\omega$:

$$I = \frac{V_{IN}}{R_{XX}} \sin\omega t \quad \text{(S1.3)}$$

The resistance shows little dependence on the input voltage and thus can be assumed to be a constant. The current induced effective fields, directed transverse to and along the current flow, are defined as $\Delta H_T \sin\omega t$ and $\Delta H_L \sin\omega t$, respectively. Here we assume that the effective field is in phase with the AC voltage excitation. The sinusoidal effective fields are included in Eq. (S1.2) to give the magnetization tilt angle $\theta(V_{IN})$ under the AC voltage application, which reads

$$\theta(V_{IN}) = \frac{\left(\left(H_T + \Delta H_T \sin\omega t\right)\cos\varphi + \left(H_L + \Delta H_L \sin\omega t\right)\sin\varphi\right)}{D} \quad \text{(S1.4)}$$

The EHE (Hall) resistance can be expressed using the magnetization tilt angle as

$$R_{XY} = \Delta R_{XY} \cos\theta \quad \text{(S1.5)}$$

where $\Delta R_{XY}$ corresponds to the difference in the EHE resistance between the two magnetization states pointing along +z and –z. Substituting Eq. (S1.4) into (S1.5) and assuming $\theta(V_{IN}) \ll 1$ (keeping terms up to $\theta^2$) gives

$$R_{XY} = \pm \Delta R_{XY}\left(1 - \frac{1}{2D^2}\left(H_T^2 \cos^2\varphi + H_L^2 \sin^2\varphi + 2H_T H_L \cos\varphi\sin\varphi\right)\right)$$
$$\mp \frac{\Delta R_{XY}}{D^2}\left(H_T \Delta H_T \cos^2\varphi + H_L \Delta H_L \sin^2\varphi + (H_T \Delta H_L + H_L \Delta H_T)\cos\varphi\sin\varphi\right)\sin\omega t + O(\Delta H_{T(L)}^2)$$
(S1.6)

Here the plus/minus sign represents the magnetization direction pointing along +z and −z, respectively. The Hall voltage, which is measured using the lock-in amplifier, is $V_{XY} = R_{XY} \cdot I$. Substituting Eqs. (S1.3) and (S1.6) into $V_{XY}$ and using the trigonometric identities, we obtain the following formula

$$V_{XY} = V_{DC} + V_\omega \sin\omega t + V_{2\omega} \cos 2\omega t$$
(S1.7)

Here $V_{DC}$ represents terms that do not depend on $\omega$. The in-phase first harmonic $V_\omega$ and the out of phase second harmonic $V_{2\omega}$ voltages are expressed as

$$V_\omega = \pm \Delta R_{XY} \frac{V_{IN}}{R_{XX}}\left(1 - \frac{1}{2D^2}\left(H_T^2 \cos^2\varphi + H_L^2 \sin^2\varphi + 2H_T H_L \cos\varphi\sin\varphi\right)\right)$$
(S1.8a)

$$V_{2\omega} = \pm \frac{\Delta R_{XY}}{2D^2} \frac{V_{IN}}{R_{XX}}\left(H_T \Delta H_T \cos^2\varphi + H_L \Delta H_L \sin^2\varphi + (H_T \Delta H_L + H_L \Delta H_T)\cos\varphi\sin\varphi\right)$$
(S1.8b)

Depending on the applied field direction, we substitute $\varphi=0$ for the in-plane transverse field and $\varphi=90$ for the longitudinal field. To obtain Eq. (1) in the main text, one needs to take the second derivative of Eq. (S1.8a) and the first derivative of Eq. (S1.8b) with respect to the external field and take their ratio.

## S2. Harmonic voltage measurements

The in-phase first harmonic Hall voltage signals, measured simultaneously with the out of phase second harmonic signals shown in Fig. 2, are shown in Fig. S1. The maximum magnetization tilt angle estimated from the first harmonic Hall voltage signal is, for example, $\theta \sim 12.6$ deg ($\theta \sim 0.23$ rad) for the data shown in Fig. S1(a). This value is within the limit of the approximation used ($\theta \ll 1$) to deduce Eq. (1). The small offset voltage in the second harmonic signals shown in Fig. 2 could be related to the small harmonic distortion of the signal generator (0.001 % total harmonic distortion). This type of offset voltage is not dependent on the applied in-plane magnetic field and thus should not affect the effective field evaluation.

## S3. Resistance versus the wedge thickness

The thickness dependence of the wire resistance, measured using the four point probe method, is shown in Fig. S2 for the Ta (S2(a)) and the CoFeB (S2(b)) wedge films. The wire resistance $R_{XX}$ is multiplied by the wire width (w: 10 μm) and divided by the length between the two voltage probes (L: 20 μm) to obtain a normalized resistance $R_{XX}\,w/L$. The slope of $1/(R_{XX}\,w/L)$ versus the wedge layer thickness gives the resistivity of the thickness varying film. We obtain $\rho$~246 μΩ·cm for the Ta layer and $\rho$~140 μΩ·cm for the CoFeB layer.

**Figure captions**

Fig. S1.  In phase first harmonic signal $V_\omega$ as a function of in-plane field directed transverse (a, c) and parallel to (b, d) to current flow for $V_{IN}$=20 $V_{pp}$.  Solid and open symbols represent signals when the magnetization is pointing along +z and –z, respectively.  The wedge layer thickness $d_{Ta}$ is 0.3 nm for (a, b) and 1.2 nm for (c, d).

Fig. S2.  Inverse of the normalized resistance as a function of the wedge layer thickness. The resistance ($R_{XX}$) measured using the four point probe method is multiplied by the wire width (w) and divided by the distance between the two voltage probes (L) to obtain the normalized resistance.

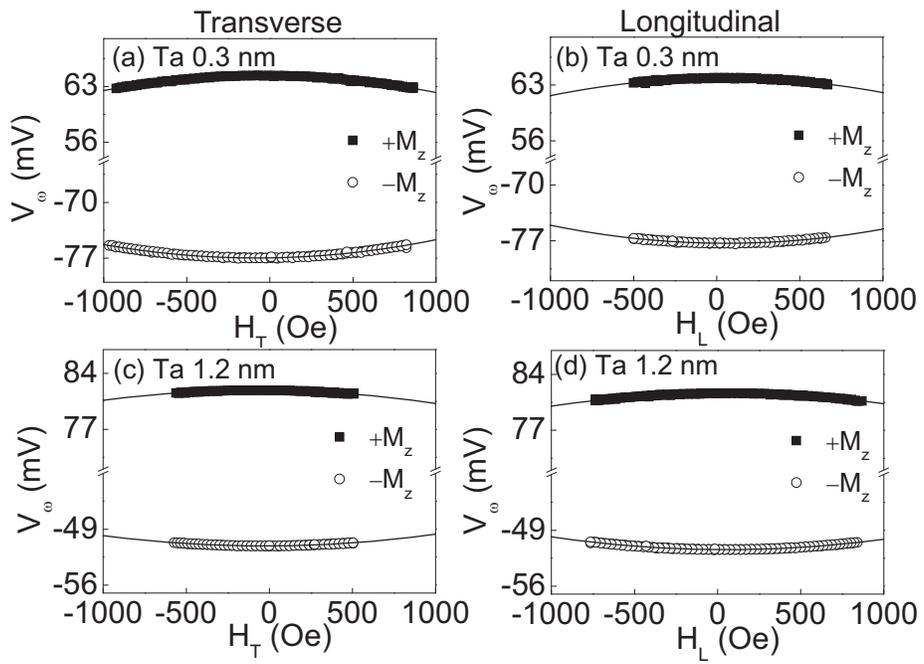

Fig. S1

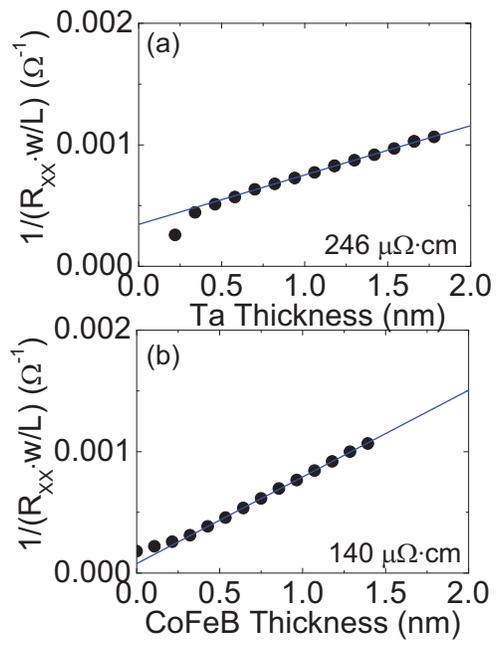

Fig. S2